\documentclass[conference] {IEEEtran}

\usepackage[utf8]{inputenc}
\usepackage[reqno]{amsmath}
\usepackage{amssymb}
\usepackage{amsthm}
\usepackage{dsfont}
\usepackage{graphics}
\usepackage{amscd}
\usepackage{enumerate}
\usepackage{graphics,xcolor,graphicx}

\usepackage{soul}
\usepackage{color}

\def\bbC{{\mathbb{C}}}

\def\bbG{{\mathbb{G}}}

\def\bf1{{\mathds{1}}}

\theoremstyle{plain}

\newtheorem{lemma}{Lemma}[section]

\newtheorem{theorem}[lemma]{Theorem}

\theoremstyle{definition}
\newtheorem{definition}[lemma]{Definition}
\newtheorem{example}[lemma]{Example}

\theoremstyle{remark}

\numberwithin{equation}{section}

\begin{document}

\title{Signal processing on large networks with group symmetries}


\author{
\IEEEauthorblockN{Kathryn {\sc Beck}, Mahya {\sc Ghandehari}, Jeannette {\sc Janssen} \& Nauzer {\sc Kalyaniwalla}}

\IEEEauthorblockA{Department of Mathematical Sciences, University of Delaware, DE, 19716, US\\
$\{${\tt kbeck,mahya$\}$@udel.edu}}

\IEEEauthorblockA{Department of Mathematics \& Statistics, Dalhousie University, NS, B3H 4R2, Canada\\
$\{${\tt jjanssen,nauzer$\}$@dal.ca}}
}
\maketitle

\section{Introduction}

Current methods of graph signal processing rely heavily on the specific structure of the underlying network: the shift operator and the graph Fourier transform are both derived directly from a specific graph. In many cases, the network is subject to error or natural changes over time. This motivated a new perspective on GSP, where the signal processing framework is developed for an entire class of graphs with similar structures (see \cite{MorencyLeus17,RuizChamonRibeiro21}). This approach can be formalized via the theory of graph limits, where graphs are considered as random samples from a distribution represented by a graphon. 

When the network under consideration has underlying symmetries, they may be modeled as samples from \emph{Cayley graphons}. In Cayley graphons, vertices are sampled from a group, and the link probability between two vertices is determined by a function of the two corresponding group elements. Infinite groups such as the 1-dimensional torus can be used to model networks with an underlying spatial reality. Cayley graphons on finite groups give rise to a \emph{Stochastic Block Model}, where the link probabilities between blocks form a (edge-weighted) Cayley graph. 


\section{Graphon signal processing}

In \cite{lovaszszegedy2006}, \emph{graphons} were introduced as limit objects of converging sequences of dense graphs.
Graphons retain the large-scale structure of the families of graphs they represent, and are hence useful representatives of these families. 

A graphon is commonly defined as a symmetric, measurable function on $[0,1]^2$, but can be defined on any standard probability space (\cite[Corollary 3.3]{Borg-Chayes-Lovasz-2010}).
In order to deal  with Cayley graphons, we adopt this approach.
\begin{definition}\label{def:graphon-signal}
Let  $(X,\mu)$ be a standard probability space, and $L^2(X)$ be the associated space of square-integrable functions. 
A function $w:X\times X\to [0,1]$ is called a graphon represented on $X$ if $w$ is measurable and symmetric (i.e.~$w(x,y)=w(y,x)$ almost everywhere). 
A graphon signal on $w$ is a pair $(w,f)$, where $f:X\rightarrow \bbC$ belongs to $L^2(X)$. 
\end{definition}

A graphon $w$ on $X^2$ can be interpreted as a probability distribution on random graphs of order $n$ denoted by ${\mathcal G}(n,w)$. Given a vertex set $\{ 1,2,\dots ,n\}$, each vertex $i$ is assigned a value $x_i$ drawn uniformly at random from $X$. Next, for each pair of vertices with labels $i<j$ independently, an edge $\{ i,j\}$ is added with probability $w(x_i,x_j)$. 
The sequence $\{{\mathcal G}(n,w)\}_n$  almost surely forms a convergent graph sequence (in the sense of graph limit theory), for which the limit object is the graphon $w$ (see \cite{lovaszszegedy2006}). 

The graphon Fourier transform is derived from the eigenspaces of the integral operator associated with the graphon.
In \cite[Theorem 3.7]{GhandehariJanssenKalyaniwalla22}, {extending results} from \cite{RuizChamonRibeiro21}, we show convergence of the Fourier transform for all graphons.
To achieve this result, we needed to redefine the graphon Fourier transform in the case where the graphon operator has eigenvalues with higher multiplicity. 
Inspired by Fourier analysis of non-Abelian groups, we replaced the concept of ``Fourier coefficients'' by projections onto eigenspaces of the shift operator. {Without going into details, our theorem says}
\begin{theorem}
If graph signals $(G_n,f_n)$ converge to a graphon signal $(w,f)$, then the vector-valued GSP for $(G_n,f_n)$ converges to that of $(w,f)$.
\end{theorem}
The following example shows that, in the presence of higher multiplicities, the individual graph Fourier coefficients do not always converge, but the projections onto eigenspaces do. 

\begin{example}\label{ex:S3_matrix}
Consider the graphon representing a stochastic block model with six equal sized blocks, and link probabilities: $p_{1,1}=0.6$, $p_{1,2}=0.3$, $p_{1,3}=0.1$, all others are zero.
(This model represents a Cayley graphon on the symmetric group $\mathbb{S}_3$.) The operator associated with the graphon has non-zero eigenvalues $\lambda_1\geq \lambda_2\geq\dots\geq\lambda_6$, where $\lambda_2=\lambda_3$. We sampled 10 graphs from the model, of order 1000 each. The first six eigenvalues of the samples were very close to those of the graphon, and the remaining eigenvalues were small.  

None of the samples had repeated eigenvalues, so they each had a unique eigenbasis. 
Taking a signal $f$ that is 1 {on the first block} and zero everywhere else,
we computed the graph Fourier coefficients of the signal on each sample. The second and third Fourier coefficients $\widehat{f}(\phi_2)=\langle f,\phi_2\rangle$ and $\widehat{f}(\phi_3)=\langle f,\phi_3\rangle$ are shown in Figure \ref{fg:projections}: each blue dot has coordinates $(\widehat{f}(\phi_3),\widehat{f}(\phi_2))$ for one sample. In the figure,  the dots do not cluster together, which shows that the graph Fourier coefficients vary greatly from sample to sample. However, as predicted by our convergence result, the vectors $\langle f,\phi_2\rangle\phi_2+\langle f,\phi_3\rangle\phi_3$ converge as the size of the graph grows; this shows since the blue dots lay close to a circle.
The red diamond shows the projection of $f$ onto the graphon Fourier basis. As predicted by the theory, this projection falls on the same circle.

\begin{figure}[ht]
\centerline{\includegraphics[scale=0.35]{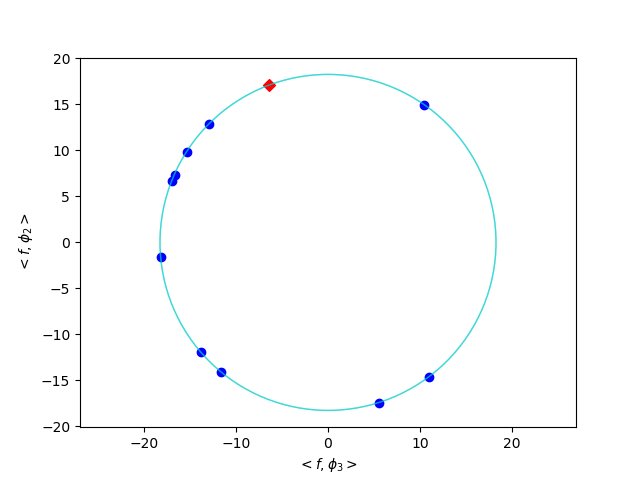}}
\caption{Graph Fourier coefficients 2 and 3 for the different samples.}
\label{fg:projections}
\end{figure}
\end{example}

\section{Cayley graphons}

Cayley graphons, first introduced in \cite{cayley-graphon}, are defined on a (finite or compact) group which, equipped with its Haar measure forms a standard probability space. The group symmetries make Cayley graphons great models for many networks.

\begin{definition}\label{def:cayley-graphon}
Let $\gamma:\bbG\rightarrow [0,1]$ be a measurable function such that $\gamma(x)=\gamma(x^{-1})$. 
Then the graphon $w:\bbG\times \bbG\to [0,1]$ defined as $w(x,y)=\gamma(xy^{-1})$ is called the Cayley graphon defined by  $\gamma$ on the group $\bbG$, and the function $\gamma$ is called a Cayley function.
\end{definition}

Note that graphs sampled from a Cayley graphon are not, in general, themselves Cayley graphs. Instead, they can be seen as ``fuzzy versions'' of Cayley graphs, which preserve the symmetries of the group on a large scale, but are locally random. 
The irreducible representations of the group underlying a Cayley graphon may be used to develop a specific framework for signal processing on the graphon, which can then be used to provide an instance-independent framework for graph signal processing.


\begin{example}[Watts-Strogatz model]\label{exp:Watts-Strogatz}
Consider the graphon $w:[0,1]^2\rightarrow [0,1]$ defined as follows. For all $x,y\in [0,1]$, let
$$
w(x,y)=\left\{ \begin{array}{ll}
1-p &\mbox{if }|x-y|\leq d\mbox{ or }|x-y|\geq 1-d,\\
p & \mbox{otherwise},
\end{array}\right.
$$
where $p,d\in (0,\frac{1}{2})$ are parameters of the model. 
The graphon $w$ is a Cayley graphon on the 1-dimensional torus. 
Random graphs drawn from $w$ have a natural circular layout: each vertex can be identified with a point on the unit circle. Then each vertex is connected with large probability $1-p$ to vertices that are close, 
and with small probability $p$ to any other vertex. 
This graphon corresponds to the Watts-Strogatz model  \cite{WattsStrogatz98}, which is widely used to model ``small-world'' networks. 
\begin{figure}[ht]
\centerline{\includegraphics[scale=0.25]{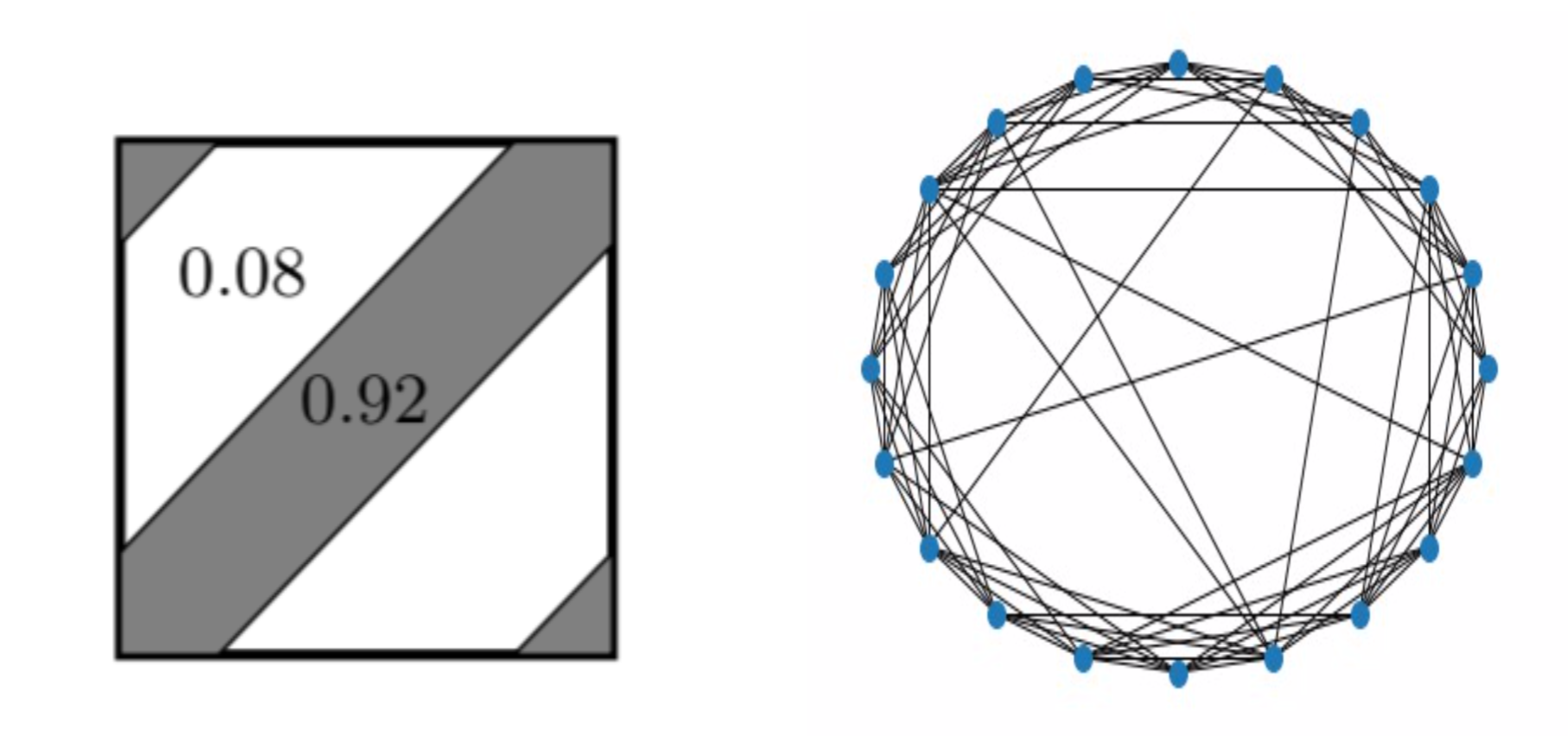}}
\caption{Cayley graphon on the 1-dim torus with parameters $d=0.2, p=0.08$ and a graph sampled from it.}
\end{figure}

\end{example}

\begin{example}[Ranking model]
Signals on Cayley graphons of the symmetric group $\mathbb S_n$ can be used to analyze ranked voting data sets. Each group element is a permutation, which represents a preference ranking of $n$ candidates.
The generating set $S$ introduces a notion of closeness between rankings. A signal on the graphon can represent the number of voters who chose that particular preference ranking. 

In \cite{permutahedron}, frames were built for the permutahedron, the Cayley graph of $\mathbb{S}_n$ with generating set of adjacent transpositions. Using these frames, the analysis coefficients of a signal provide interpretation of the ranked data set, such as popularity or polarity of candidates. This meaningful frame construction was extended in \cite{BeckGhandehari23} to the setting of any Cayley graph.

As an example, consider the Cayley graph of $\mathbb S_4$ with generating set $S=\{(12), (23), (34),(12)(34)\}$. In the ranking interpretation, this means that permutations are considered close if {either two adjacent candidates are switched, or if both the first two and the last two candidates switch orders.}

The frame construction starts by listing all irreducible unitary representations of $\mathbb S_4$.
 A frame can be obtained from the eigenvalues and eigenvectors for the matrices $\pi(S)=\sum_{s\in S}\pi(s)$ where $\pi$ is an irreducible representation. In all but one case, $\pi(S)$ has no repeated eigenvalues, meaning there is only one choice of a nontrivial frame. In the remaining case, 
 we can build a Mercedes-Benz frame for the eigenspace, 
 which is known for its simplicity and for being a tight frame.
 A recipe is then used to lift each of these frames for the {low-dimensional} spaces to a frame for $\ell^2(\mathbb S_4)$,
 which is suitable for signal processing on the Cayley graph.

\end{example}


\end{document}